\documentclass[12pt,a4paper,final]{iopart}
\usepackage{iopams}

\expandafter\let\csname equation*\endcsname\relax
\expandafter\let\csname endequation*\endcsname\relax

\usepackage{amsmath}
\usepackage{amsfonts}
\usepackage{amsmath}
\usepackage{amssymb}
\usepackage{amsthm}
\usepackage[utf8]{inputenc}
\usepackage{bm}
\usepackage{graphicx}
\usepackage{cite}
\usepackage{comment}

\usepackage[breaklinks=true,colorlinks=true,linkcolor=blue,urlcolor=blue,citecolor=blue]{hyperref}

\newcommand{\defeq}{\mathrel{\mathop:}=}

\newcommand{\D}{\mathcal{D}}
\renewcommand{\H}{\mathcal{H}}

\begin{document}

\title{Configurational density of states of finite classical systems}

\author[cor1]{Sergio Davis$^{1,2}$, Boris Maulén$^2$}
\address{$^1$Research Center on the Intersection in Plasma Physics, Matter and Complexity (P$^2$mc), Comisión Chilena de Energía Nuclear, Casilla 188-D, Santiago, Chile}
\address{$^2$Departamento de Física y Astronomía, Facultad de Ciencias Exactas, Universidad Andres Bello, Sazié 2212, piso 7, 8370136, Santiago, Chile}
\ead{sergio.davis@cchen.cl}

\begin{abstract}
The configurational density of states (CDOS) encodes all the relevant thermodynamic information contained in the interaction potentials for statistical mechanical systems. However, its explicit computation is usually a challenge
for non-trivial systems, and numerical algorithms such as Wang-Landau simulation are often used. In this work we use a microcanonical framework to provide an explicit inversion formula for the calculation of the CDOS from the 
total density of states (DOS) without resorting to the inversion of the Laplace transform. From this formula, several results can be obtained for the thermodynamics of finite classical systems composed of a few degrees of freedom, 
while also recovering the well-known asymptotic results for the thermodynamic limit.
\end{abstract}

\section{Introduction}

For an interacting system of particles with interaction potential $\Phi(\bm R)$, the configurational density of states (CDOS), defined by
\begin{equation}
\D(\phi) \defeq \int d\bm{R}\,\delta\big(\Phi(\bm R)-\phi)
\end{equation}
evaluates the density of configurational microstates with energies within $\phi$ and $\phi+\mathrm{d}\phi$, and contains all the information from the interaction potential $\Phi$ that is relevant to the thermodynamical 
properties of the system. Actual computation of the CDOS, however, is a challenge for most interaction potentials, including the widely used Lennard-Jonnes potential~\cite{lennard-jonnes}, the Morse and Kratzer potentials for 
interatomic interactions within a molecule~\cite{kratzer}, among many others. Monte Carlo methods such as Wang-Landau simulation~\cite{Wang2001,Moreno2022} and methods based on configurational temperature~\cite{Calvo1995,Rathore2003,Rathore2003a,Escobedo2006} or Bayesian inference~\cite{Moreno2023b} allow for the determination of the CDOS but are sometimes limited by its high computational cost.


In this work, we show that for a system of $N$ classical particles, precise knowledge of the total density of states (DOS), defined by
\begin{equation}
\Omega(E) \defeq \int d\bm{P}d\bm{R}\,\delta\left(\sum_{i=1}^N \frac{\bm{p}_i^2}{2m_i} + \Phi(\bm R)-E\right)
\end{equation}
is sufficient to recover the CDOS. We present an explicit formula that derives the CDOS from the total DOS for finite $N$ in a purely microcanonical framework. This is achieved by solving the generalized Abel's integral 
equation (AIE)\cite{Polyanin2008, gakhov1966}, which belongs to the Volterra class of integral equations, i.e., those in which the integral upper limit is not a constant value, but is a variable one. Additionally, the AIE 
admits a full analytical solution, whereby the solution for the CDOS shown in this work is exact for all $N \geq 1$. It is important to highlight that we have worked within the classical microcanonical ensemble, which makes 
it possible to obtain a unique probability density associated with kinetic and potential energies that are compatible with a given total DOS.

This article is organized as follows. In Section \ref{total_config_dos}, we review both the total and the configurational density of states in the framework of the microcanonical ensemble, showing their connection through an 
integral transform. In Section \ref{inversion_formula_sect} we state the inversion formula from the total DOS to the CDOS, while in Section \ref{proof} we give a proof of the inversion formula stated previously, by transforming 
our integral equation into a generalized Abel's integral equation. As a first application of our inversion formula, in Section \ref{sec:constant_CE} we explicitly compute the CDOS for a system with constant heat capacity. 
One of the most important results obtained from our approach is presented in Section \ref{concave_region}, where we study a system with non-constant heat capacity where the DOS exhibits a concave region. In this case we obtain 
its CDOS from the knowledge of the corresponding total DOS. The importance of these kind of features of the DOS is that it allows the description of metastable states in a first-order phase transition~\cite{kubo_book,Schnabel2011}. 
Finally, in Section \ref{constat_heat_capacity} we highlight the usefulness of our approach by deriving the microcanonical distribution of particle velocities for a system with constant heat capacity, without the need to invert 
the Laplace transform as in the canonical approach. In this context, we obtain a single-particle velocity distribution which is different from the Maxwell-Boltzmann distribution, whereas the multi-particle distribution converges 
to the Maxwellian in the thermodynamic limit $N\rightarrow \infty$. In addition, we include an Appendix in which we show the resolution of the generalized Abel's integral equation in detail.

\section{Total and configurational density of states} \label{total_config_dos}

Consider a classical system of $N$ particles with Hamiltonian
\begin{equation}
\label{eq:ham}
\H(\bm R, \bm P) = K(\bm P) + \Phi(\bm R),
\end{equation}
where $K(\bm P)$ is the kinetic energy term, given as usual by
\begin{equation}
K(\bm P) \defeq \sum_{i=1}^N \frac{\bm{p}_i^2}{2 m_i},
\end{equation}
and $\Phi(\bm R)$ is the potential energy term, including arbitrary interactions but only dependent on the configurational degrees of freedom $\bm R$. 
If the system is isolated, having fixed total energy $E$, its distribution of microstates will be described by the microcanonical ensemble,
\begin{equation}
P(\bm R, \bm P|E) = \frac{\delta\big(\H(\bm R, \bm P)-E\big)}{\Omega(E)},
\end{equation}
where $\Omega(E)$ is the total density of states, given by
\begin{equation}
\label{eq:dos}
\Omega(E) \defeq \int d\bm{R}d\bm{P}\,\delta\big(\H(\bm R, \bm P)-E\big).
\end{equation}

\noindent
We can use the $N$-particle kinetic density of states
\begin{equation}
\label{eq:kdos_N}
\Omega_K(K; N) \defeq \int d\bm{p}_1\ldots d\bm{p}_N\,\delta\left(\sum_{i=1}^N \frac{\bm{p}_i^2}{2 m_i} -K\right)
\end{equation}
in order to perform the integration on $\bm P$. Upon the change of variables
\begin{equation}
\bm{u}_i \defeq \frac{\bm{p}_i}{\sqrt{2 m_i K}}
\end{equation}
we can write \eqref{eq:kdos_N} as
\begin{equation}
\Omega_K(K; N) = W_N \big[K\big]_+^{\frac{3N}{2}-1}
\end{equation}
where the notation $\big[X\big]_+ = X\Theta(X)$ with $\Theta$ the Heaviside step function, that is, $\big[X\big]_+$ evaluates to $X$ for $X \geq 0$, and zero otherwise. The constant $W_N$ is given by
\begin{equation}
W_N \defeq \prod_{i=1}^N (2 m_i)^{\frac{3}{2}}\times\left[\int d\bm{u}_1\ldots d\bm{u}_N \delta\left(\sum_{i=1}^N \bm{u}_i^2 - r^2\right)\right]_{r = 1}
\end{equation}
and is such that the integral in brackets can be obtained as
\begin{equation}
\int d\bm{u}_1\ldots d\bm{u}_N \delta\left(\sum_{i=1}^N \bm{u}_i^2 - r^2\right) = \frac{1}{2r}\frac{\partial}{\partial r}\int d\bm{u}_1\ldots d\bm{u}_N
\Theta\left(r^2 - \sum_{i=1}^n \bm{u}_i^2\right) = \frac{1}{2r}\frac{\partial V_{3N}(r)}{\partial r}
\end{equation}
where
\begin{equation}
V_{3N}(r) = \frac{2\,(\pi)^{\frac{3N}{2}} r^{3N}}{3N\Gamma\left(\frac{3N}{2}\right)}
\end{equation}
is the volume of the $3N$-dimensional hypersphere~\cite{Weisstein2002} of radius $r$. Therefore we finally have
\begin{equation}
W_N \defeq \frac{1}{\Gamma\big(\frac{3N}{2}\big)}\prod_{i=1}^N\big(2\pi m_i\big)^\frac{3}{2},
\end{equation}
and we can rewrite \eqref{eq:dos} as
\begin{equation}
\label{eq:dos_pre}
\Omega(E) = \int d\bm{R}\left[\int d\bm{P}\,\delta\big(K(\bm P)-E+\Phi(\bm R)\big)\right] = W_N\int d\bm{R}\Big[E-\Phi(\bm R)\Big]_+^{\frac{3N}{2}-1}.
\end{equation}

\noindent
At this point, we introduce the configurational density of states, defined by
\begin{equation}
\D(\phi) \defeq \int d\bm{R}\,\delta\big(\Phi(\bm R)-\phi\big),
\end{equation}
and such that, for any function $G(\phi)$ we have
\begin{equation}
\label{eq:cdos_prop}
\int d\bm{R}\,G(\Phi(\bm R)) = \int_0^\infty d\phi\,\D(\phi)G(\phi).
\end{equation}

\noindent
Using the property \eqref{eq:cdos_prop} on the rightmost integral in \eqref{eq:dos_pre}, and defining $\tilde{D}(\phi)$ as
\begin{equation}
\tilde{\D}(\phi)=\prod_{i=1}^N \left( 2\pi m_i\right)^{3/2}  \D(\phi)
\end{equation}
we finally have
\begin{equation}
\Omega(E)=\left( \mathbb{I}^{\alpha} \tilde{\D} \right)(E), \label{eq:omega}
\end{equation}
where $\mathbb{I}^{\alpha}$ is the \textit{Riemann-Liouville fractional integral} of order $\alpha=3N/2$, which is defined by \cite{f_calculus}
\begin{equation}
\left( \mathbb{I}^{\alpha} \tilde{\D} \right)(E) \defeq \dfrac{1}{\Gamma(\alpha)} \int_{0}^E d\phi \hspace{0.07cm} \D(\phi) \left( E-\phi\right)^{\alpha-1}.
\end{equation}
Hence, Eq. (\ref{eq:omega}) gives the total density of states $\Omega(E)$ as a fractional integral transform of $\tilde{\D}(\phi)$.
In the next section we will show that, given $\Omega(E)$ for a particular system, the integral equation in \eqref{eq:omega} can be solved for $\D(\phi)$.

\section{An inversion formula} \label{inversion_formula_sect}

We can prove that, for a given $\Omega(E)$ under certain conditions such as the existence of its derivatives up to the $3(N+1)/2$-th order, the corresponding $\D(\phi)$ is unique, and is formally given by
\begin{equation}
\D(\phi)=\left( \mathbb{D}^{\lambda} f \right)(\phi) \defeq \dfrac{1}{\Gamma(1-\lambda)} \dfrac{d}{d\phi} \int_0^{\phi} d\varepsilon \hspace{0.1cm} \dfrac{f(\varepsilon)}{(\phi-\varepsilon)^{\lambda}}, \label{eq:formal_solution}
\end{equation}
where $\mathbb{D}^{\lambda}$ is the \textit{Riemann-Liouville fractional derivative} of order $\lambda$ \cite{f_calculus}. In Eq. (\ref{eq:formal_solution}) the Riemann-Liouville fractional derivative is acting on a function $f$ which is built from the $(m+1)$-th derivative of the DOS, i.e.
\begin{equation}
f(\varepsilon) \defeq \dfrac{1}{\Gamma\big(\frac{3N}{2}\big)} \dfrac{1}{\Gamma(1-\lambda)} \dfrac{1}{W_N} \Omega^{(m+1)}(\varepsilon)
\end{equation}
with $\lambda$ defined as
\begin{equation}
\lambda \defeq \frac{1}{2}\big(N\;\text{mod}\;2\big) = \begin{cases}
0\;\text{for}\;\text{even}\;N, \\[5pt]
\frac{1}{2}\;\text{for}\;\text{odd}\;N,
\end{cases}
\end{equation}
and
\begin{equation}
m \defeq \frac{3N}{2}-\lambda-1.
\end{equation}

%
%
%
\section{Proof of the inversion formula} \label{proof}

Differentiating \eqref{eq:omega} with respect to $E$ on both sides, we have
\begin{equation}
\left(\frac{3N}{2}-1\right)\int_0^\infty d\phi \hspace{0.1cm} \D(\phi)\Theta(E-\phi)(E-\phi)^{\frac{3N}{2}-2} = \frac{1}{W_N}\frac{\partial \Omega(E)}{\partial E},
\end{equation}
and, in general, the $m$-th derivative of \eqref{eq:omega} gives
\begin{equation}
\prod_{k=1}^m \left(\frac{3N}{2}-k\right)\int_0^\infty d\phi \hspace{0.1cm} \D(\phi)\Theta(E-\phi)(E-\phi)^{\frac{3N}{2}-1-m} = \frac{1}{W_N}\hspace{0.1cm}\Omega^{(m)}(E)
\end{equation}
so the integral on the left-hand side is
\begin{equation}
\label{eq:integral}
\int_0^E d\phi \hspace{0.1cm} \D(\phi)(E-\phi)^{\frac{3N}{2}-1-m} = \frac{\Gamma\big(\frac{3N}{2}-m\big)}{\Gamma\big(\frac{3N}{2}\big)}\,\frac{1}{W_N}\hspace{0.1cm} \Omega^{(m)}(E).
\end{equation}

\noindent
Now, the value of $m$ can be chosen so that the exponent
\begin{equation}
\lambda \defeq \frac{3N}{2}-m-1 \label{lambda}
\end{equation}
is such that $0 < \lambda < 1$. This can be acomplished by the choice
\begin{equation}
m = \begin{cases}
\frac{3N}{2}-1\quad\text{if}\,\,N\,\text{is even},\\[10pt]
\frac{3(N-1)}{2}\quad\text{if}\,\,N\,\text{is odd},
\end{cases}
\end{equation}
such that we have $\lambda = 0$ for even $N$, and $\lambda = 1/2$ for odd $N$. We arrive at the integral equation
\begin{equation}
\int_0^E d\phi \hspace{0.1cm} \D(\phi)(E-\phi)^\lambda 
= \frac{\Gamma(\lambda+1)}{\Gamma\big(\frac{3N}{2}\big)}\,\frac{1}{W_N}\hspace{0.1cm}\Omega^{(m)}(E), \label{integral_eq}
\end{equation}
which can be solved for $\D(\phi)$. In order to solve \eqref{integral_eq}, we recast it by differentiating on both sides with respect to the energy $E$, thus obtaining
\begin{equation}
\int_{0}^{E} d\phi \hspace{0.1cm} \dfrac{\D(\phi)}{(E-\phi)^{1-\lambda}}=\dfrac{\Gamma(\lambda)}{\Gamma\big(\frac{3N}{2}\big)} \dfrac{1}{W_N} \hspace{0.1cm} \Omega^{(m+1)}(E). \label{integral_eq}
\end{equation}

The integral equation in \eqref{integral_eq} is a generalized form of the \textit{Abel's integral equation} \cite{Polyanin2008, gakhov1966} (AIE), whose resolution is shown in detail in the Appendix. In particular, 
using the solution given in \eqref{solution_abel} with $\mu=1-\lambda$, we obtain that the configurational density of states $\D(\phi)$ is of the form
\begin{equation}
\D(\phi)=\dfrac{1}{\Gamma\big(\frac{3N}{2}\big)\Gamma(1-\lambda)} \dfrac{1}{W_N} \dfrac{d}{d\phi} \int_0^\phi d\varepsilon\hspace{0.1cm} \dfrac{\Omega^{(m+1)}(\varepsilon)}{(\phi-\varepsilon)^{\lambda}}.
\end{equation}

\section{The case of constant microcanonical heat capacity}
\label{sec:constant_CE}

One of the most, if not the most, important families of thermodynamical systems is that with constant microcanonical heat capacity. This includes the ideal gas and the harmonic oscillator, among several others, including 
most solids for temperatures well below their melting point. In this case the microcanonical caloric curve, that is, temperature as a function of total energy, is linear, as in
\begin{equation}
E = \alpha k_B T_E,
\end{equation}
hence, the microcanonical heat capacity $C_E$ is constant, and given by
\begin{equation}
C_E \defeq \left(\frac{\partial E}{\partial T}\right)_E = \alpha k_B,
\end{equation}
where $\alpha$ is dimensionless and proportional to the number of particles in the thermodynamic limit (i.e. it is extensive). The microcanonical inverse temperature, defined by
\begin{equation}
\label{eq:beta_om}
\frac{1}{k_B T_E} = \beta_\Omega(E) = \frac{\partial}{\partial E}\ln \Omega(E),
\end{equation}
is then
\begin{equation}
\beta_\Omega(E) = \frac{\alpha}{E},
\end{equation}
and by integration we obtain the total density of states as
\begin{equation}
\label{eq:omega_ex}
\Omega(E) = \Omega_0 E^\alpha,
\end{equation}
with $\Omega_0$ an integration constant. Upon replacing the $(m+1)$-th derivative of \eqref{eq:omega_ex},
\begin{equation}
\Omega^{(m)}(\varepsilon) = \frac{\Gamma(\alpha+1)}{\Gamma(\alpha-m)}\Omega_0\,E^{\alpha-m-1} 
\end{equation}
into \eqref{eq:formal_solution} we have
\begin{equation}
\D(\phi) = \frac{1}{\Gamma\big(\frac{3N}{2}\big)\Gamma(1-\lambda)} \hspace{0.1cm} \frac{\Omega_0}{W_N} 
\frac{\Gamma(\alpha+1)}{\Gamma(\alpha-m)}\frac{d}{d \phi}\int_0^\phi d\varepsilon \hspace{0.1cm} \frac{\varepsilon^{\alpha-m-1}}{(\phi-\varepsilon)^\lambda}. \label{microcanonical1}
\end{equation}
The integral in \eqref{microcanonical1} can be easily computed with the aid of the beta function. The result is
\begin{equation}
\int_0^\phi d\varepsilon \hspace{0.1cm} \frac{\varepsilon^{\alpha-m-1}}{(\phi-\varepsilon)^\lambda} = B(1-\lambda,\alpha-m)\hspace{0.1cm} \phi^{\alpha-m-\lambda}
= \frac{\Gamma(1-\lambda)\Gamma(\alpha-m)}{\Gamma(\alpha+1-m-\lambda)} \hspace{0.1cm} \phi^{\alpha-m-\lambda}. \label{microcanonical2}
\end{equation}
Then, after performing the derivative with respect to $\phi$ on \eqref{microcanonical2}, using the definition of $\lambda$ given in \eqref{lambda} and simplifying, we finally obtain for a system with constant microcanonical heat capacity $C_E = \alpha k_B$ and arbitrary size $N \geq 1$, that its configurational density of states is
\begin{equation}
\label{eq:cdos_alpha}
\D(\phi) = D_0\,\phi^{\alpha-\frac{3N}{2}},
\end{equation}
where the constant $D_0$ is given by
\begin{equation}
\label{microcanonical4}
D_0 \defeq \frac{\Omega_0}{W_N}\frac{\Gamma(\alpha+1)}{\Gamma\big(\frac{3N}{2}\big)\Gamma(\alpha+1-\frac{3N}{2})}.
\end{equation}

A common assumption, mostly in short-range interacting systems, is that the energy $E$ is extensive, that is, $E = \varepsilon N$. If we also assume that the temperature is intensive (independent of $N$), 
then it follows that $\alpha = aN$ with $c_E = a k_B$ the heat capacity per atom.

\section{Density of states with a concave region} \label{concave_region}

As another example of the application of our main result in \eqref{eq:formal_solution}, we will now consider the density of states
\begin{equation}
\label{eq:dos_inflect}
\Omega(E) = \Omega_0\big(1 + b^2(E-\mu)^2\big)E^\alpha
\end{equation}
with $\alpha > 0$, $b \geq 0$ and $\mu > 0$, such that the microcanonical heat capacity is no longer constant for $b \neq 0$. The microcanonical inverse temperature $\beta_\Omega$, defined in 
\eqref{eq:beta_om}, is for this model given by
\begin{equation}
\beta_\Omega(E) = \frac{\partial}{\partial E}\ln \Omega(E) = \frac{\alpha}{E} + \frac{2b^2 (E-\mu)}{1 + b^2(E-\mu)^2}.
\end{equation}

For $b\mu > \sqrt{\alpha(\alpha+2)}$, the density of states $\Omega(E)$ in \eqref{eq:dos_inflect} has two inflection points, at $E = E_m \pm \Delta$, where
\begin{align}
E_m & \defeq \frac{\mu(\alpha+1)}{\alpha+2}, \\
\Delta & \defeq \frac{\sqrt{(b\mu)^2 - \alpha(\alpha+2)}}{b(\alpha+2)},
\end{align}
such that $\beta_\Omega(E) < 0$ for $\big|E-E_m\big| < \Delta$. This kind of region is usually associated with metastable states in a first-order phase transition~\cite{Schnabel2011}, such that the 
canonical energy distribution is bimodal, as shown for instance in Ref.~\cite{Junghans2008}.

We can readily compute the CDOS associated to \eqref{eq:dos_inflect} by writing it as a linear combination of constant heat capacity terms, namely
\begin{equation}
\Omega(E) = \Omega_0\big(1 + (b\mu)^2\big)E^\alpha + \Omega_0(b^2)E^{\alpha+2} - 2\Omega_0(b^2)\mu E^{\alpha+1}.
\end{equation}
 
Due to the linearity of the derivatives and integral in \eqref{eq:formal_solution}, we can directly apply the result in \eqref{eq:cdos_alpha} term by term, obtaining
\begin{equation}
\D(\phi) = \frac{\Omega_0 \Gamma(\alpha+1)}{W_N\,\Gamma\left(\frac{3N}{2}\right)\Gamma\left(\alpha+1-\frac{3N}{2}\right)}G(\phi; \alpha, b, \mu, N)\phi^{\alpha-\frac{3N}{2}}
\end{equation}
where we have defined the function
\begin{equation}
G(\phi; \alpha, b, \mu, N) = 1 + (b\mu)^2 + \frac{b^2(\alpha+1)(\alpha+2)\phi^2}{\left(\alpha+1-\frac{3N}{2}\right)\left(\alpha+2-\frac{3N}{2}\right)} - \frac{2(b^2)\mu\,(\alpha+1)\phi}{\alpha+1-\frac{3N}{2}}.
\end{equation}
for convenience. Taking $\alpha = aN$ we obtain in the the thermodynamic limit that
\begin{equation}
\lim_{N \rightarrow \infty} G(\phi; \alpha, b, \mu, N) = 1 + \frac{b^2}{\left(1-\frac{3}{2a}\right)^2}\left[\phi - \left(1-\frac{3}{2a}\right)\mu\right]^2,
\end{equation}
then $\D(\phi)$ takes a similar form to $\Omega(E)$, with new parameters
\begin{align}
b' & \defeq \frac{2ab}{2a - 3}, \\
\mu' & \defeq \frac{\mu(2a-3)}{2a}.
\end{align}

\section{Microcanonical velocity distribution for constant heat capacity} \label{constat_heat_capacity}

The microcanonical distribution of configurations is
\begin{equation}
P(\bm R|E) = \frac{\Omega_K\big(E-\Phi(\bm R); N\big)}{\Omega(E)} = \frac{W_N}{\Omega(E)}\Big[E-\Phi(\bm R)\Big]_+^{\frac{3N}{2}-1}
\end{equation}
so the corresponding distribution of potential energies is given by
\begin{equation}
\label{eq:probphi}
P(\phi|E) = \int d\bm{R}\,P(\bm R|E)\delta\big(\Phi(\bm R)-\phi\big) = \frac{W_N}{\Omega(E)}\Big[E-\phi\Big]_+^{\frac{3N}{2}-1}\D(\phi).
\end{equation}

\noindent
This kind of expression corresponds to a $q$-canonical distribution, with
\begin{equation}
q \defeq 1 + \frac{2}{2 - 3N}
\end{equation}
as noted by Naudts and Baeten~\cite{Naudts2009}. Because $\Omega(E)$ determines $\D(\phi)$, the microcanonical potential energy distribution in \eqref{eq:probphi} is completely determined by $\Omega(E)$ alone. 
The same is true for the microcanonical distribution of single-particle velocities, as we have
\begin{equation}
P(\bm{p}_1, \ldots, \bm{p}_N|E) = \frac{1}{\Omega(E)}\int d\bm{R}\,\delta\left(\frac{\bm{p}_1^2}{2m_1} + \sum_{i=2}^N \frac{\bm{p}_i^2}{2 m_i}+\Phi(\bm R)-E\right).
\end{equation}

The single-particle momentum distribution is obtained by integrating out the remaining momenta,
\begin{equation}
P(\bm{p}_1|E) = \int d\bm{p}_2\ldots d\bm{p}_N\;P(\bm{p}_1, \ldots, \bm{p}_N|E),
\end{equation}
hence, calling $\bm{p}_1 \rightarrow \bm{p}$ and $m_1 \rightarrow m$, we have
\begin{equation}
P(\bm{p}|E) = \frac{1}{\Omega(E)}\int d\bm{p}_2\ldots d\bm{p}_N\int d\bm{R}\,\delta\left(E-\Phi(\bm R)-\frac{\bm{p}^2}{2 m} - \sum_{i=2}^N \frac{\bm{p}_i^2}{2 m_i}\right).
\end{equation}
By using \eqref{eq:cdos_prop} to introduce the CDOS, the single-particle momentum distribution simplifies to
\begin{equation}
\begin{split}
P(\bm{p}|E) & = \frac{1}{\Omega(E)}\int d\bm{p}_2\ldots d\bm{p}_N\int d\phi\,\D(\phi)\,\delta\left(E-\phi-\frac{\bm{p}^2}{2 m} - \sum_{i=2}^N \frac{\bm{p}_i^2}{2 m_i}\right) \\
& = \frac{1}{\Omega(E)}\int_0^\infty d\tilde{K}\int d\phi\,\D(\phi)\,\delta\big(E-k(\bm v)-\phi-\tilde{K}\big)\Omega_K(\tilde{K}; N-1)
\end{split}
\end{equation}
where we have replaced $\bm{v} = \bm{p}/m$ and defined
\begin{equation}
k(\bm v) \defeq \frac{m\bm{v}^2}{2}
\end{equation}
to be the single-particle kinetic energy. Making the change of variables from $\bm p$ to $\bm{v} = \bm{p}/m$, we can write the single-particle velocity distribution as
\begin{equation}
P(\bm v|E) = m\,P(\bm{p} = m\bm v|E),
\end{equation}
and replacing \eqref{eq:kdos_N}, finally leads to
\begin{equation}
P(\bm v|E) = \frac{m\,W_{N-1}}{\Omega(E)}\int_0^{E-k(\bm v)}d\phi\,\big[E-k(\bm v)-\phi\big]_+^{\frac{3(N-1)}{2}-1}\D(\phi).
\end{equation}

\begin{figure}[htb]
\centering
\includegraphics[width=0.7\linewidth]{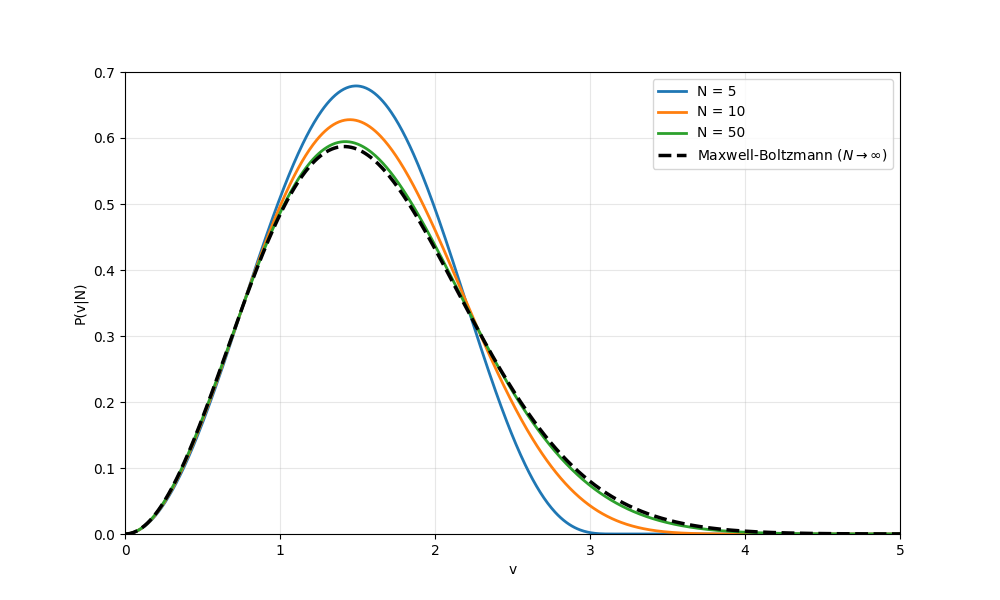}
\caption{Comparison between q-distributions of molecular speeds for a finite N with the Maxwell-Boltzmann distribution (thermodynamic limit).} \label{fig1}
\end{figure}

The distribution $P(\bm v|E)$ is, in general, different from the Maxwell-Boltzmann distribution for finite $N$, but converges to it in the thermodynamic limit. In the case developed in 
Section~\ref{sec:constant_CE}, we have
\begin{equation}
\label{eq:veldist}
P(\bm v|E) = \left(\frac{m}{2\pi}\right)^\frac{3}{2}\frac{\Gamma(\alpha+1)}{E^\alpha\Gamma(\alpha-1/2)}\Big[E-\frac{m\bm{v}^2}{2}\Big]_+^{\alpha-\frac{3}{2}},
\end{equation}
a formula previously obtained by Ray and Graben~\cite{Ray1991b} by inversion of the Laplace transform. This expression can, in principle, be used for the precise determination of the heat capacity in 
atomistic computer simulations, from collected samples of atomic velocities. By taking $E = \varepsilon N$ and $\alpha = a N$, we can rewrite \eqref{eq:veldist} as
\begin{equation}
P(\bm v|\varepsilon, N) = \left(\frac{m}{2\pi\,\varepsilon}\right)^\frac{3}{2}N^{-\frac{3}{2}}\frac{\Gamma(aN+1)}{\Gamma(aN-1/2)}\left[1-\left(\frac{m\bm{v}^2}{2\varepsilon N}\right)\right]_+^{aN-\frac{3}{2}}
\end{equation}
which is a $q$-Gaussian distribution~\cite{Tsallis2009c} with
\begin{equation}
\label{eq:qvalue}
q = 1 - \frac{2}{2aN-3} < 1.
\end{equation}
The fact that $q < 1$ in \eqref{eq:qvalue} makes the model in \eqref{eq:veldist} \emph{subcanonical}~\cite{Davis2022b}, therefore incompatible with the superstatistical formalism~\cite{Beck2003}. 
In the thermodynamic limit, $N \rightarrow \infty$, so $q \rightarrow 1$ and we recover the Maxwell-Boltzmann distribution
\begin{equation}
\lim_{N \rightarrow \infty} P(\bm v|\varepsilon, N) = \left(\frac{m}{2\pi k_B T_E}\right)^\frac{3}{2}\exp\Big(-\frac{m\bm{v}^2}{2 k_B T_E}\Big)
\end{equation}
with
\begin{equation}
k_B T_E = \frac{E}{\alpha} = \frac{\varepsilon}{a}.
\end{equation}

The mean kinetic energy per particle is proportional to the temperature, in a finite-size version of the equipartition theorem,
\begin{equation}
\big<k\big>_E = \frac{3E}{2(\alpha+1)} = \frac{3}{2}\left(\frac{\alpha}{\alpha+1}\right)k_B T_E \leq \frac{3}{2}k_B T_E,
\end{equation}
where equality is reached in the limit $\alpha \rightarrow \infty$. On the other hand, the relative variance of the kinetic energy per particle is
\begin{equation}
\frac{\big<(\delta k)^2\big>_E}{\big<k\big>_E^2} = \frac{2}{3}-\frac{5}{3(\alpha+2)} \leq \frac{2}{3},
\end{equation}
agreeing in the limit $\alpha \rightarrow \infty$ with the value of $2/3$ for the Maxwell-Boltzmann distribution, as expected.

\section{Concluding remarks}

We have presented the inversion formula \eqref{eq:formal_solution}, that gives the configurational density of states (CDOS) in terms of the total density of states (DOS). This result clearly shows that the CDOS is unique 
for a given total DOS, while also providing an explicit, exact expression by means of which the CDOS can be computed for finite systems and in the thermodynamic limit. This exact expression may provide some insights on the 
features of first-order phase transitions in finite systems, as well as enabling the computation of thermodynamic properties of complex systems where one only has access to the DOS.

\section*{Acknowledgments}

SD and BM gratefully acknowledge funding from ANID FONDECYT 1220651 grant.

%
%
\newpage
\renewcommand{\theequation}{A-\arabic{equation}}
  \setcounter{equation}{0}  

\section*{Appendix: Generalized Abel's integral equation} \label{Appendix}
The generalized Abel's integral equation is given by
\begin{equation}
    \int_{a}^t dt' \dfrac{y(t')}{(t-t')^{\mu}}=g(t). \label{appendix1}
\end{equation}
In order to solve \eqref{appendix1}, we integrate between $a$ and $x$ after multiplying on both sides by $(x-t)^{\mu-1}$, i.e.
\begin{equation}
\int_{a}^x dt \hspace*{0.1cm} (x-t)^{\mu-1} \int_{a}^t dt' \dfrac{y(t')}{(t-t')^{\mu}}  =\int_{a}^x dt \hspace{0.1cm}  \dfrac{g(t)}{(x-t)^{1-\mu}}. \label{appendix2}
\end{equation}
Consider the left side of \eqref{appendix2}. By use of the Dirichlet's formula for reversal of the integration order, namely
\begin{equation}
\int_{a}^x dt \int_{a}^t dt' \hspace{0.1cm} F(x,t,t')=\int_{a}^x dt' \int_{t'}^x dt \hspace{0.1cm} F(x,t,t'), 
\end{equation}  
we can recast the left side of \eqref{appendix2} in the following way
\begin{equation}
\int_{a}^x dt  \int_{a}^t dt' (x-t)^{\mu-1} (t-t')^{-\mu} \hspace{0.1cm} y(t')   =\int_{a}^x dt' \left\lbrace \int_{t'}^x dt \hspace{0.1cm}(x-t)^{\mu-1} (t-t')^{-\mu} \right\rbrace y(t'). \label{curly_brackets}
\end{equation}

The advantage of the previous reversal of the integration order is that we can compute the integral within the curly brackets in (\ref{curly_brackets}) independently. In this sense, after defining a new variable $s$ as
\begin{equation}
s = \dfrac{t-t'}{x-t'}
\end{equation}
we obtain
\begin{equation}
\int_{t'}^x dt \hspace{0.1cm}(x-t)^{\mu-1} (t-t')^{-\mu} = \int_{0}^1 ds \hspace{0.1cm} s^{-\mu} (1-s)^{\mu-1}=B(1-\mu,\mu),
\end{equation}
where $B(1-\mu,\mu)$ is the beta function, which is related with the gamma function by
\begin{equation}
B(1-\mu,\mu)=\dfrac{\Gamma(1-\mu) \Gamma(\mu)}{\Gamma(1-\mu+\mu)}=\dfrac{\Gamma(1-\mu) \Gamma(\mu)}{\Gamma(1)}=\Gamma(1-\mu) \Gamma(\mu).
\end{equation}
Hence, the integral within the curly brackets in \eqref{curly_brackets} becomes
\begin{equation}
\int_{t'}^x dt \hspace{0.1cm}(x-t)^{\mu-1} (t-t')^{-\mu} = \Gamma(1-\mu) \Gamma(\mu).
\end{equation}
Using this result in the integral equation (\ref{appendix2}),
\begin{equation}
\int_a^x dt' \hspace{0.1cm} y(t') = \dfrac{1}{\Gamma(1-\mu) \Gamma(\mu)} \int_a^x dt \hspace{0.1cm} \dfrac{g(t)}{(x-t)^{1-\mu}},
\end{equation}
and invoking the fundamental theorem of calculus, we finally obtain the desired solution of the generalized Abel's integral equation written in terms of the Riemann-Liouville fractional derivative:
\begin{equation}
\label{solution_abel}
y(x) = \dfrac{1}{\Gamma(1-\mu) \Gamma(\mu)} \hspace{0.1cm} \dfrac{d}{dx} \int_a^x dt \hspace{0.1cm} \dfrac{g(t)}{(x-t)^{1-\mu}}=\left( \mathbb{D}^{\lambda} \tilde{g} \right)(x), 
\end{equation}
where we have taken $\lambda=1-\mu$ and $\tilde{g}(x) \defeq g(x)/\Gamma(\lambda)$.
\section*{References}

\bibliography{cdos}
\bibliographystyle{unsrt}

\end{document}